\documentclass[]{spie}  %>>> use for US letter paper
\usepackage{amsmath,amsfonts,amssymb}
\usepackage[skip=0pt]{subcaption}
\usepackage[skip=5pt]{caption}
\usepackage{graphicx}
\usepackage[colorlinks=true, allcolors=blue]{hyperref}
\usepackage{orcidlink}

\title{The Simons Observatory: Dark Characterization of the Large Aperture Telescope}

\author[1]{Saianeesh K.	Haridas\orcidlink{0000-0001-6519-502X}}
\author[4,5]{Zeeshan Ahmed\orcidlink{0000-0002-9957-448X}}
\author[1]{Tanay Bhandarkar\orcidlink{0000-0002-2971-1776}}
\author[1]{Mark Devlin\orcidlink{0000-0002-3169-9761}}
\author[1]{Simon Dicker\orcidlink{0000-0002-1940-4289}}
\author[10]{Shannon M. Duff\orcidlink{0000-0002-9693-4478}}
\author[2]{Daniel Dutcher\orcidlink{0000-0002-9962-2058}}
\author[11,3]{Kathleen Harrington\orcidlink{0000-0003-1248-9563}}
\author[4,5]{Shawn W. Henderson\orcidlink{0000-0001-7878-4229}}
\author[10]{Johannes Hubmayr\orcidlink{0000-0002-2781-9302}}
\author[9]{Bradley R. Johnson\orcidlink{0000-0002-6898-8938}}
\author[1]{Anna Kofman\orcidlink{0000-0001-5374-1767}}
\author[1]{Alex Manduca}
\author[7,8]{Michael D. Niemack\orcidlink{0000-0001-7125-3580}}
\author[12]{Michael J. Randall\orcidlink{0009-0009-9806-2317}}
\author[6,4]{Thomas P. Satterthwaite\orcidlink{0000-0002-6452-4220}}
\author[1]{John Orlowski-Scherer\orcidlink{0000-0003-1842-8104}}
\author[1]{Benjamin L. Schmitt}
\author[3]{Carlos Sierra}
\author[13,14]{Max Silva-Feaver\orcidlink{0000-0001-7480-4341}}
\author[1]{Robert J. Thornton}
\author[2]{Yuhan Wang\orcidlink{0000-0002-8710-0914}}
\author[2]{Kaiwen Zheng\orcidlink{0000-0003-4645-7084}}

\affil[1]{Department of Physics and Astronomy, University of Pennsylvania, Philadelphia, PA 19146; USA}
\affil[2]{Joseph Henry Laboratories of Physics, Princeton University, Princeton, NJ 08542; USA}
\affil[3]{Department of Physics, University of Chicago, Chicago, IL 60637; USA}
\affil[4]{Kavli Institute for Particle Astrophysics and Cosmology, Stanford, CA 94305; USA}
\affil[5]{SLAC National Accelerator Laboratory, Menlo Park, CA 94025; USA}
\affil[6]{Department of Physics, Stanford University, Stanford, CA 94305; USA}
\affil[7]{Department of Physics, Cornell University, Ithaca, NY 14853; USA}
\affil[8]{Department of Astronomy, Cornell University, Ithaca, NY 14853; USA}
\affil[9]{Department of Astronomy, University of Virginia, Charlottesville, VA 22904; USA}
\affil[10]{Quantum Sensors Division, National Institute of Standards and Technology, Boulder, CO 80305; USA}
\affil[11]{Argonne National Laboratory, High Energy Physics Division, Lemont, IL 60439; USA}
\affil[12]{Department of Physics, University of California, San Diego, La Jolla, CA, 92093; USA}
\affil[13]{Department of Physics, Yale University, New Haven, CT 06511; USA}
\affil[14]{Wright Laboratory, Yale University, New Haven, CT 06511; USA}

\authorinfo{Corresponding author: Saianeesh K. Haridas\\ E-mail: haridas@sas.upenn.edu}

\begin{document} 
\maketitle

\begin{abstract}
The Simons Observatory (SO) is a cosmic microwave background experiment composed of three 0.42 m Small Aperture Telescopes (SATs) and one 6 m Large Aperture Telescope (LAT) in the Atacama Desert of Chile.
The Large Aperture Telescope Receiver (LATR) was integrated into the LAT in August 2023; however, because mirrors were not yet installed, the LATR optical chain was capped at the 4K stage.
In this dark configuration we are able to characterize many elements of the instrument without contributions from atmospheric noise.
Here we show this noise is below the required upper limit and its features are well described with a simple noise model.
Maps produced using this noise model have properties that are in good agreement with the white noise levels of our dark data.
Additionally, we show that our nominal scan strategy has a minimal effect on the noise when compared to the noise when the telescope is stationary.

\end{abstract}

% Include a list of keywords after the abstract 

\section{Introduction}
The Simons Observatory (SO)\cite{SO_Goals} is a cosmic microwave background (CMB) experiment currently undergoing deployment and commissioning at a high elevation (5200 m) site in the Atacama Desert of Chile.
SO consists of three 0.42m Small Aperture Telescopes (SATs)\cite{galitzki2024simons} and one 6m Large Aperture Telescope (LAT)\cite{Zhu_2021}.
The LAT is a crossed-Dragone telescope\cite{Parshley_2018}, giving it a large focal plane that enables a high mapping speed.
The LAT's detectors are housed in the Large Aperture Telescope Receiver (LATR)\cite{Xu_2021}, which contains $\sim$30,000 transition edge sensor (TES) bolometers.
These TESs are distributed across 21 Universal Focal plane Modules (UFMs)\cite{Healy_2022} that are spread across 7 discrete optics tubes (OTs).
Each OT contains detectors with two frequencies that are grouped into three bands: low-frequency (LF) containing 30 and 40 GHz, mid-frequency (MF) containing 90 and 150 GHz\cite{mccarrick202190}, and ultra high-frequency (UHF) containing 230 and 280 GHz\cite{Healy_2022}.
Within the UFMs the detectors are grouped in pixels containing four detectors with two frequencies and two polarizations, these pixels are rotated relative to each other giving us a total of twelve polarizations.
In the future, the LATR will be upgraded to contain 6 more OTs with the Advanced Simons Observatory (ASO)\cite{ASO}, bringing the detector count up to $\sim$62,000.

\begin{figure}[h]
    \centering
    \begin{subfigure}{.45\textwidth}
        \includegraphics[width=\textwidth]{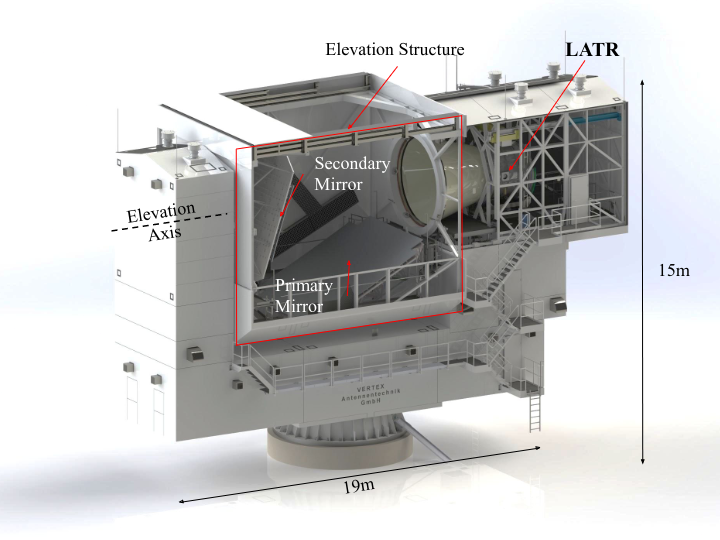}
    \end{subfigure}
    \begin{subfigure}{.45\textwidth}
        \includegraphics[width=\textwidth]{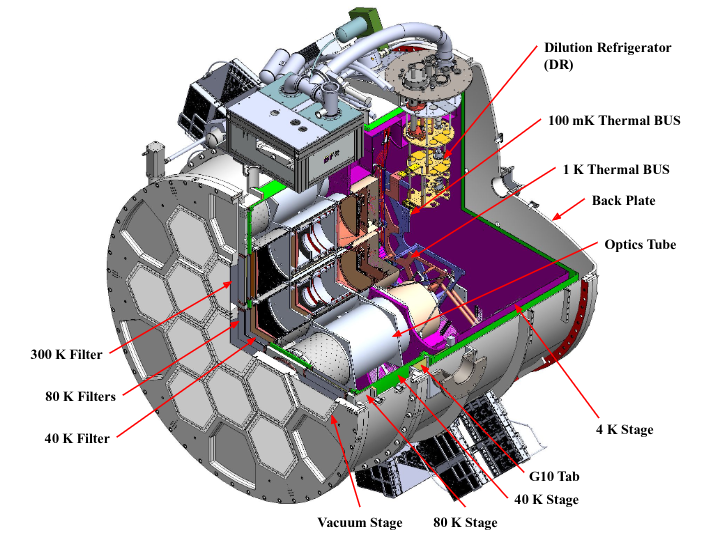}
    \end{subfigure}
    \caption{The LAT and LATR. Figures from \citenum{Xu_2021}.}
\end{figure}

\subsection{Dark Configuration}
\begin{figure}[h]
    \centering
    \begin{subfigure}{.6\textwidth}
        \includegraphics[width=\textwidth]{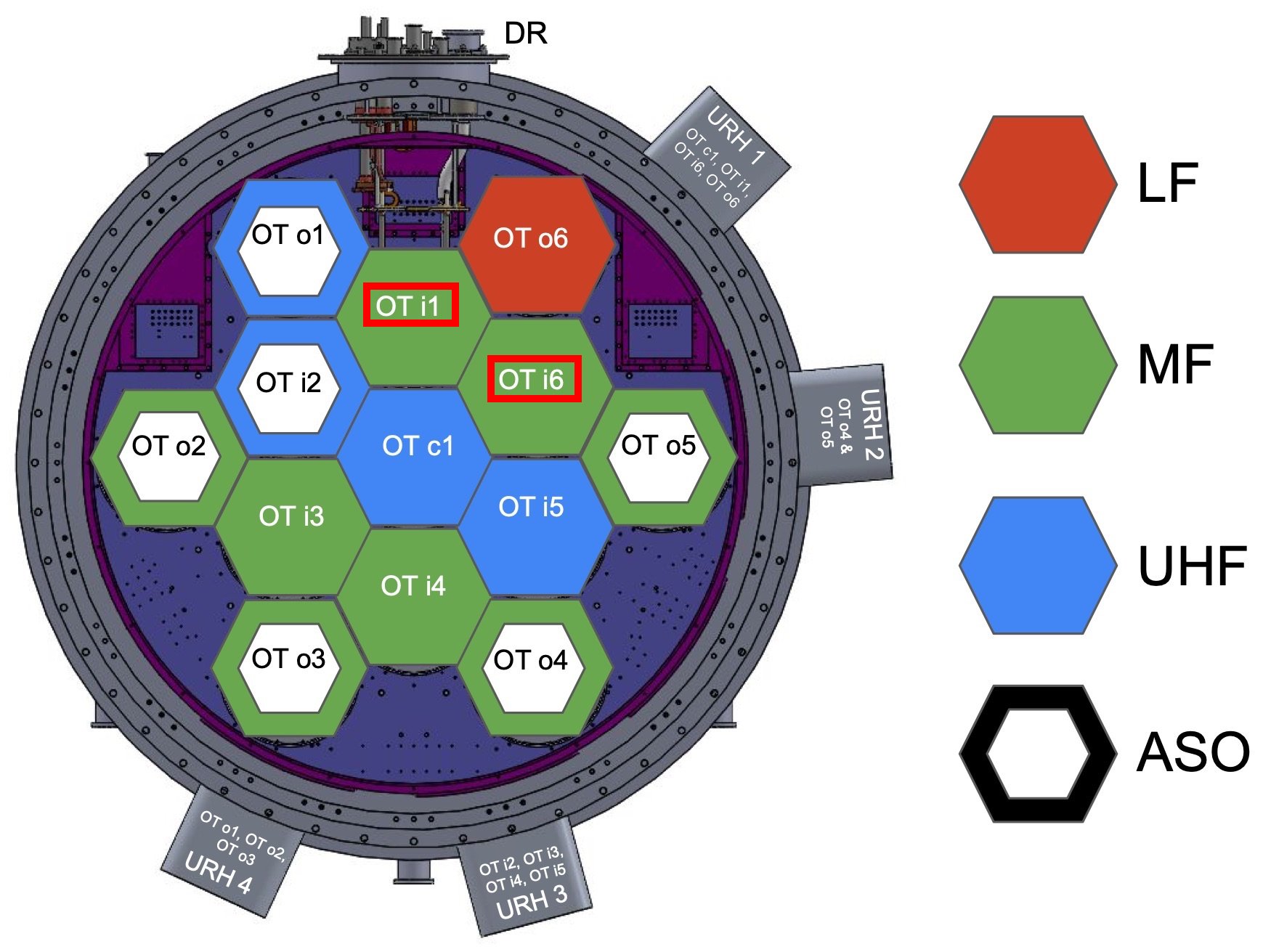}
    \end{subfigure}
    \begin{subfigure}{.35\textwidth}
        \includegraphics[width=\textwidth]{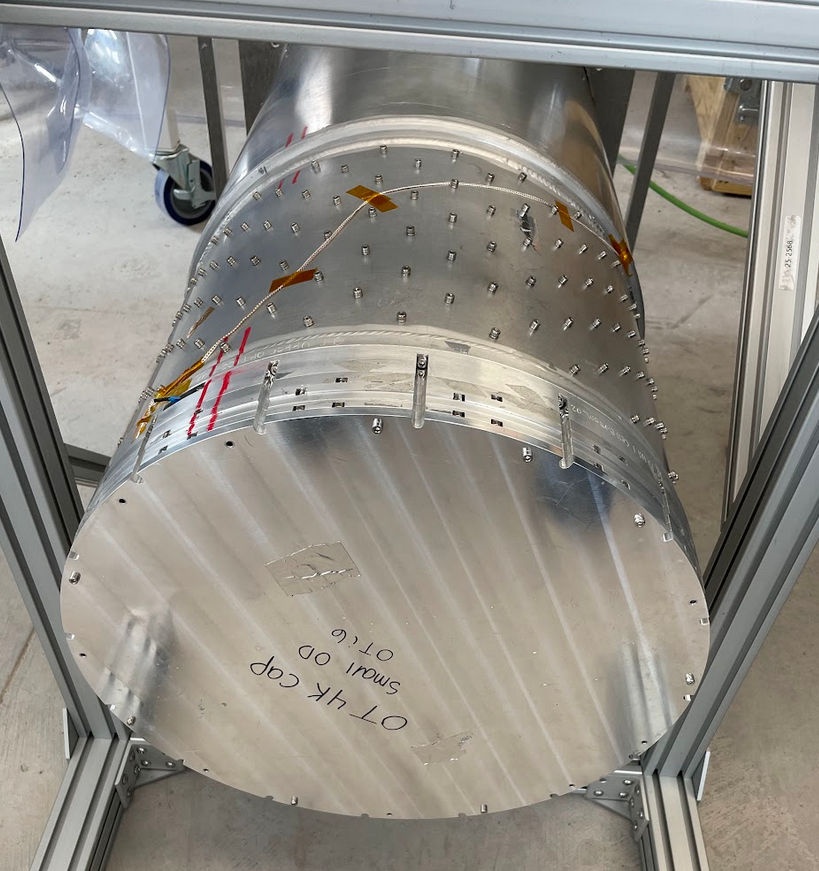}
    \end{subfigure}
    
    \caption{OT and 4K cap locations. OT i1 and i6 (boxed in red) were installed with a cap on the end of the 4K stage (pictured on the right) in order to limit the loading incident on the detectors. The resulting dark noise measurements are an important check on the detector and readout performance of the LATR.}
    \label{fig:ots}
\end{figure}

In 2023, the LATR was integrated into the LAT with two mid-frequency OTs (6 UFMs) installed.
While the LATR was able to cool down and take data, the LAT mirrors were still under construction, preventing us from taking on-sky data.
Were we to leave the LATR open optically without mirrors, our detectors would have saturated since the surfaces in the LAT are at a much higher temperature than the sky.
In order to allow us to characterize the system ahead of the mirrors arriving, we opted to install aluminum plates at the end of the two OTs we had installed at the time, capping them at the 4K stage and exposing the detectors to much lower loading (see Figure \ref{fig:ots}).
All the data presented here was taken in this dark configuration.
Subsequently, more OTs have been installed; preliminary characterization is reported in \citenum{Toby}, but detailed characterization as presented here are forthcoming. 

\section{Time Ordered Data Processing} \label{tod_proc}

\begin{figure}[h]
    \centering
    \begin{subfigure}{.45\textwidth}
        \includegraphics[trim={0 0 0 1.1cm},clip,width=\textwidth]{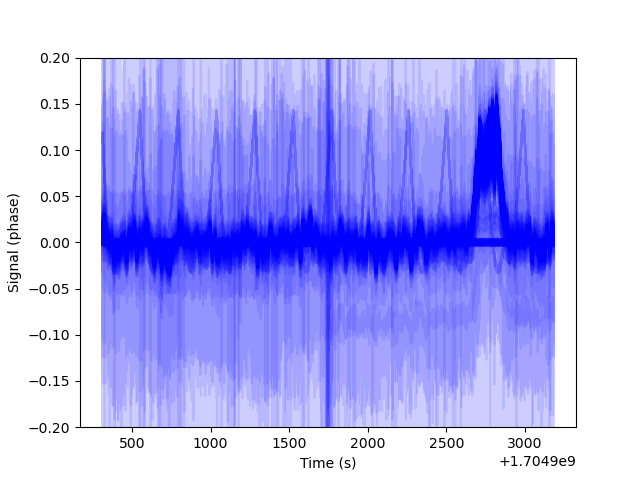}
        \caption{Before processing}
    \end{subfigure}
    \begin{subfigure}{.45\textwidth}
        \includegraphics[trim={0 0 0 1.1cm},clip,width=\textwidth]{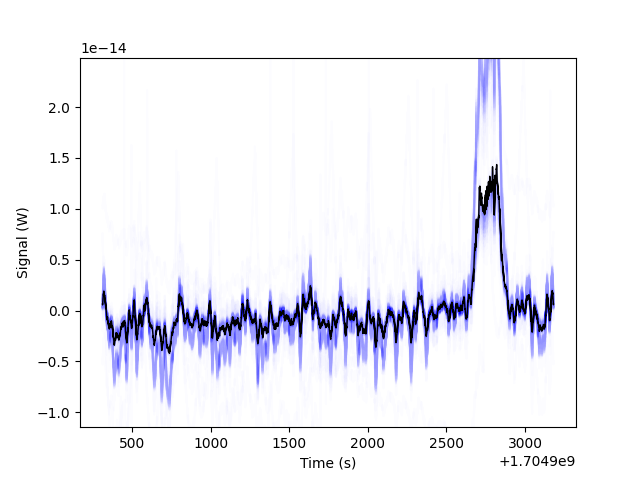}
        \caption{After processing}
        \label{fig:tod_flagged}
    \end{subfigure}
    \caption{A typical dark TOD before and after our processing. Here we have masked out glitches and repaired jumps. Detectors that are unlocked, over flagged ($>10\%$ of samples flagged), or that have high white noise (more than 3 times the median) are removed. This results in $\sim15\%$ of detectors being cut. Work to improve our pipeline to reduce the number of cut detectors is ongoing. After cutting we calibrate the TOD to Kelvin as described in Section \ref{tod_cal}. The visible structure in the TOD is dominated by small changes in the bath temperature (see Figure \ref{fig:temp}). The blue lines are $\sim20\%$ of the detectors in a single UFM ($\mathcal{O}(100)$ individual detectors).}
    \label{fig:tod_raw_to_flagged}
\end{figure}

The detectors in the LATR are read out using the SLAC Microresonator Radio Frequency (SMuRF) readout system.
Each detector is coupled to a resonator whose frequency is tracked with SMuRF, and a change in the power on the detector results in a shift in the resonator; see [\citenum{2023RScI...94a4712Y}] for a more detailed treatment.
SMuRF samples these resonator shifts at 4 kHz and subsequently down-samples them to 200 Hz time-ordered data (TODs).
Before using these data to characterize the instrument, we must process them.
In particular, there are a number of flagging steps required to detect and address problematic features in our data, either by removing or repairing them (see Figure \ref{fig:tod_raw_to_flagged}).
Additionally, correlated noise is much easier to characterize in the time domain, so it is beneficial to make an estimate of the noise with the TODs prior to mapmaking.
This is true for filter-and-bin style maps \cite{Schaffer_2011}, such as the ones presented in Section \ref{mapmaking}, where we filter the TODs before directly binning them into pixels to form a map; as well as maximum likelihood mapmakers\cite{D_nner_2012}, which will be used by the LAT once the mirrors are installed.
While the processing presented here was successful in providing us with data useful for this dark characterization, work is still ongoing to improve the efficiency and accuracy of our techniques.

\subsection{Flagging} \label{tod_flag}

Raw TODs will contain a number of unwanted phenomena that must be dealt with prior to any data reduction steps.
Here we discuss three phenomena that are commonly seen in LAT data, though this is not an exhaustive list.
Much of SO's flagging inherits techniques from the Atacama Cosmology Telescope (ACT) \cite{D_nner_2012},
but significant development effort was required to adapt these techniques for use with SO.
It should also be noted that after all the processing steps here, we apply additional cuts for detectors with abnormally high white noise values or peak-to-peak swings, but those remove $<1\%$ of the remaining detectors.
We also cut detectors that contain a large proportion of flagged samples, this is typically the largest source of cuts.
After flagging the typical TOD has 10-30\% of its detectors cut, current work is ongoing to understand this high flagging rate.
A preliminary investigation of this high flagging rate on a small number of TODs show that our false positive rate when flagging is relatively high, work to improve this is ongoing.

\begin{figure}[h]
    \centering
    \begin{subfigure}{.3\textwidth}
        \includegraphics[trim={0 0 0 1.3cm},clip,width=\textwidth]{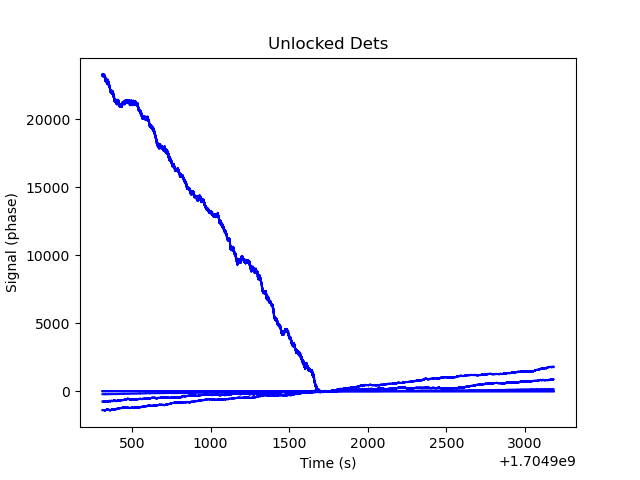} 
        \caption{Unlocked detectors}
        \label{fig:example_unlock}
    \end{subfigure}
    \begin{subfigure}{.3\textwidth}
        \includegraphics[trim={0 0 0 1.3cm},clip,width=\textwidth]{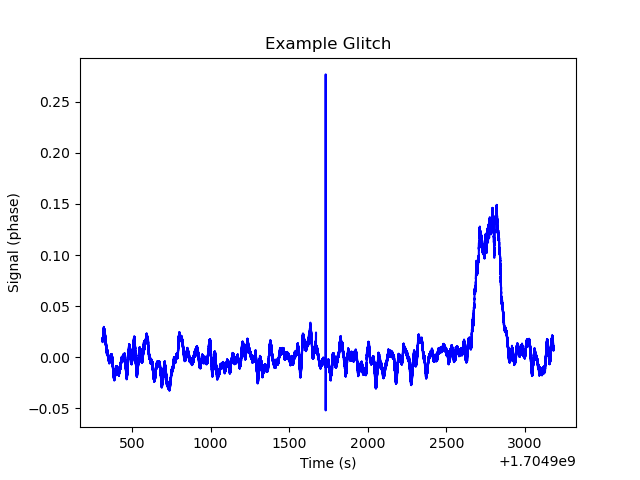} 
        \caption{TOD with a glitch}
        \label{fig:example_glitch}
    \end{subfigure}
    \begin{subfigure}{.3\textwidth}
        \includegraphics[trim={0 0 0 1.3cm},clip,width=\textwidth]{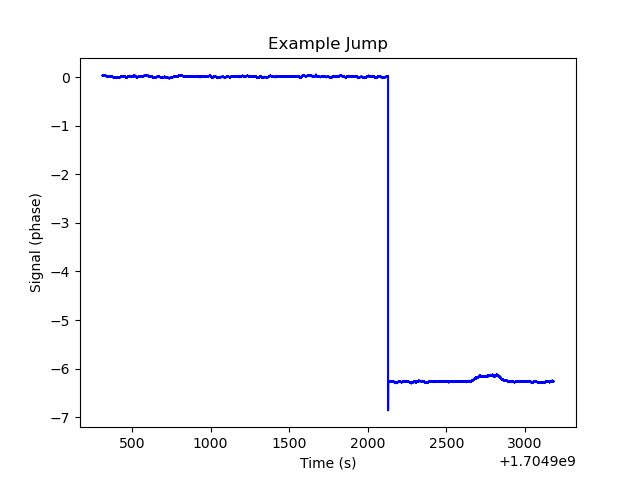} 
        \caption{TOD with a jump}
        \label{fig:example_jump}
    \end{subfigure}
    \caption{Examples of problematic phenomena that are flagged as described in Section \ref{tod_flag}.
    Unlocked detectors are completely cut. Glitches are masked and replaced using a PCA model. Jumps are subtracted and then the location of the jump is treated as a glitch.}
\end{figure}

\subsubsection{Unlocks}
If the SMuRF electronics lose the ability to track a resonator, then the associated readout channel is now ``unlocked".
While these channels can be recovered, the recovery process only happens periodically (once a day in the dark data).
Any TODs from an unlocked channel are not meaningful, so identifying and removing them from our dataset is key.
Unlocked channels have TODs that change with a slope that is a factor of $\sim 1000$ higher than what one expects in a normal observation (see Figure \ref{fig:example_unlock}), so they can be identified by comparing the least squares slope against a threshold that was determined based on lab characterization of the instrument.
This ramping is caused by the feedback loop becoming unstable when a channel is unlocked \cite{2023RScI...94a4712Y}.

\subsubsection{Glitches}
Glitches are sudden spikes that appear in TODs that only affect a small number of samples (see Figure \ref{fig:example_glitch}),
often caused by cosmic rays.
Glitches will bias the pixel in the map they are binned into to a more extreme value than the true temperature of the sky in that pixel.
Additionally, sharp features like glitches tend to ring in Fourier space, which is problematic since Fourier filters are commonly used in data reduction pipelines (such as the mapmaking pipeline implemented for ACT \cite{Naess_2020}).
To find these glitches, we look for spikes in the absolute value of TODs that have been high-pass filtered using a sine-squared filter and gaussian smoothed.
Once found, the glitches are replaced with data approximated from a principal component analysis (PCA) model of the TOD.

\subsubsection{Jumps}
Jumps are a step function shift of the mean level in the TOD  (see Figure \ref{fig:example_jump}).
A feature is only considered a jump if this shift occurs over the course of order 1 ($\mathcal{O}(1)$) samples,
making it too fast and sharp to be caused by fluctuations in the bath temperature or sky signal.
The sources of SO's jumps are still under investigation, but two suspected causes are issues in the readout and radio frequency interference.
Since jumps have a characteristic shape that is well modeled by a step function, we employ a matched filter to identify the jumps.
Once found, jumps are repaired by subtracting off their estimated height.
Since there is some uncertainty in the location of the jump, the samples around it are treated the same as a glitch and replaced using PCA.

\subsection{Common Mode Estimation} \label{cm}
\begin{figure}[h]
    \centering
    %\begin{subfigure}{.3\textwidth}
    %    \includegraphics[trim={0 0 0 1.1cm},clip,width=\textwidth]{figures/tod_proc/tods.png}
    %    \caption{Before common-mode subtraction}
    %    \label{fig:tod_precm}
    %\end{subfigure}
    \begin{subfigure}{.45\textwidth}
        \includegraphics[trim={0 0 0 1.1cm},clip,width=\textwidth]{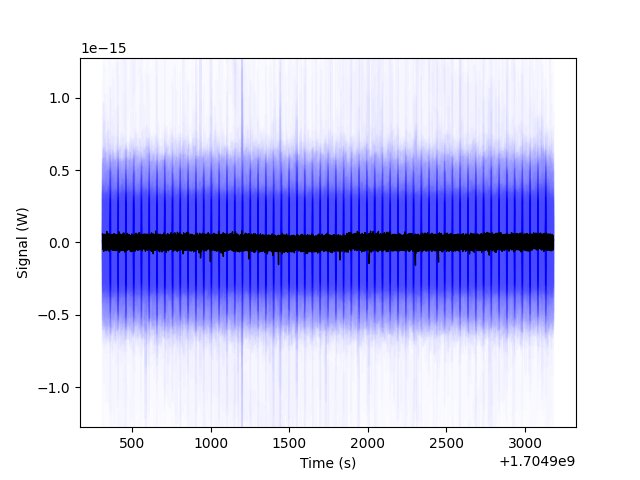}
        \caption{After common-mode subtraction}
        \label{fig:tod_cmsub_only}
    \end{subfigure}
    \begin{subfigure}{.45\textwidth}
        \includegraphics[trim={0 0 0 1.1cm},clip,width=\textwidth]{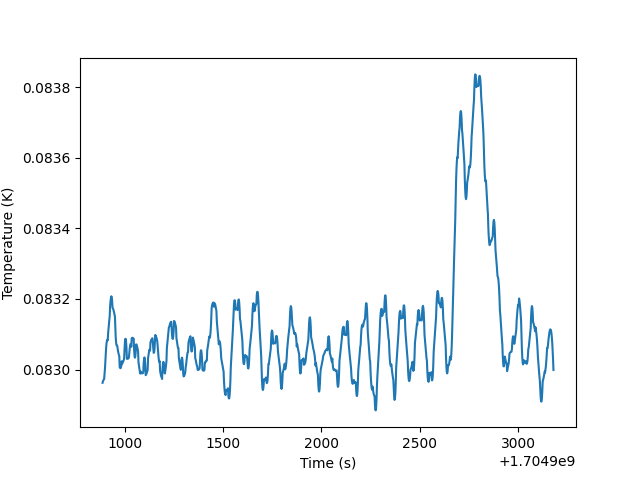}
        \caption{UFM temperature}
        \label{fig:temp}
    \end{subfigure}
    \caption{The flagged TOD pictured in Figure \ref{fig:tod_flagged} after common-mode subtraction along with the UFM temperature at the same time. The blue lines in the first plot are $\sim20\%$ of the detectors in a single UFM ($\mathcal{O}(100)$ individual detectors) that have been flagged as described in Section \ref{tod_flag} and the black lines are the medians of the pictured detectors. The second plot is the temperature of the UFM at the same time as the TOD with a third order Savitzky-Golay filter applied to reduce noise from the thermometry readout. Note that much of the structure in the TOD prior to the common mode subtraction (Figure \ref{fig:tod_flagged}) is similar to the structure in the UFM temperature and that after common mode subtraction the TOD lacks any strong features or trends.}
    \label{fig:tod_cmsub}
\end{figure}

In the dark configuration, we expect the TOD signal will be dominated by noise sourced by fluctuations in the bath temperature.
We expect to be similarly noise-dominated once on-sky, except with the atmosphere being the primary source \cite{D_nner_2012}, so modeling the noise in noise-dominated TODs will be key for producing good maps of the CMB.
Since all the detectors in a given UFM experience the same bath fluctuations, we can approximate the noise as the common mode of the TOD.
We model this common mode by using the singular value decomposition (SVD) to obtain the eigen-decomposition of the TOD.
Any variability for different detectors within a given UFM in the coupling to the bath will be captured by each detector's eigenvalues.
Additionally, correlations only between subsets of detectors can be captured as different eigenmodes, so any localized effects (e.g., variations in the atmosphere across the focal plane on-sky) are also well modeled by the SVD.
Due to memory constraints, it is unfeasible to keep all $N_{det}$ modes produced by the SVD for our common mode model, so we instead restrict ourselves to the top 50 modes\footnote{Note that this number was chosen entirely due to practical limits on shared computing resources; work to determine how many modes are required to produce a good common mode is ongoing.}.
Since we are using a reduced number of modes, this common mode can also be thought of as a low rank approximation of the TOD.
As seen in Figure \ref{fig:tod_cmsub}, subtracting off the top 50 modes yields a TOD that lacks any obvious trends or features; the impact of this on our noise is explored in Section \ref{tod_noise}.
Note that in Figure \ref{fig:tod_cmsub_only} there is structure above the noise level; these originate from detectors with high noise and spurious features that are not well described by the common mode.
Prior to mapmaking, we employ an additional cut for detectors with high noise after the common-mode subtraction to filter these out; this is a very small ($\mathcal{O}(0.1\%)$) number of detectors.

\subsection{Calibration to Power} \label{tod_cal}
The TODs output by SMuRF are in units of phase, which lack physical meaning and are difficult to interpret.
We can convert this to a sky temperature by using
\begin{equation}
    d_{K} = d_{\phi}\frac{dI}{d\phi}\frac{dP}{dI}\frac{dT}{dP}
    \label{eq:cal}
\end{equation}
where $d_{K}$ is the data in sky temperature, $d_{\phi}$ is the data is phase, $\frac{dI}{d\phi}$ is the conversion from phase to current, $\frac{dP}{dI}$ is the conversion from current to power, and $\frac{dT}{dP}$ is the conversion from power to temperature.

Normally $\frac{dT}{dP}$ is computed from measurements of sources on-sky, which we are unable to perform in the dark configuration.
So we do not include it in the work and instead convert the data to power, $d_{P}$, by removing the $\frac{dT}{dP}$ term in Equation \ref{eq:cal}.
The conversion from phase to current, $\frac{dI}{d\phi}$ is determined by the mutual inductance of the superconducting quantum interference devices (SQUIDs) \cite{760596} used in the UFMs \cite{Healy_2022}\cite{mccarrick202190} with the TESes.
This value was measured to be $\frac{9 \times10^{6} pA}{2\pi}$\cite{Dober_2021}.
The conversion from current to power, $\frac{dP}{dI}$, is done by adding a small square wave to the DC bias level of the detectors.
A detailed treatment of how this is used to compute this conversion can be found in \citenum{Wang_2022}.

\section{TOD Noise} \label{tod_noise}

Once the TODs are processed, we can use them to measure noise in the dark configuration.
When on-sky, the addition of the atmospheric noise will make it more difficult to characterize any small features in instrument noise,
so it is beneficial to understand them ahead of time using the dark noise.

In this section we model the TOD noise $n$ in Fourier space as
\begin{equation}
    n = n_{w}\left( 1 + {\left(\frac{f_{knee}}{f}\right)}^{\alpha}\right) \;,
    \label{eq:1f_noise}
\end{equation}
where $n_{w}$ is the white noise level, $f$ is the frequency, $f_{knee}$ is the knee of the $\frac{1}{f}$ noise, and $\alpha$ is the index of the $\frac{1}{f}$ noise.
The amplitude spectral densities (ASDs) of the data presented here were computed using Welch's method\cite{1161901}.

\begin{figure}[h]
    \centering
    \includegraphics[scale=.8]{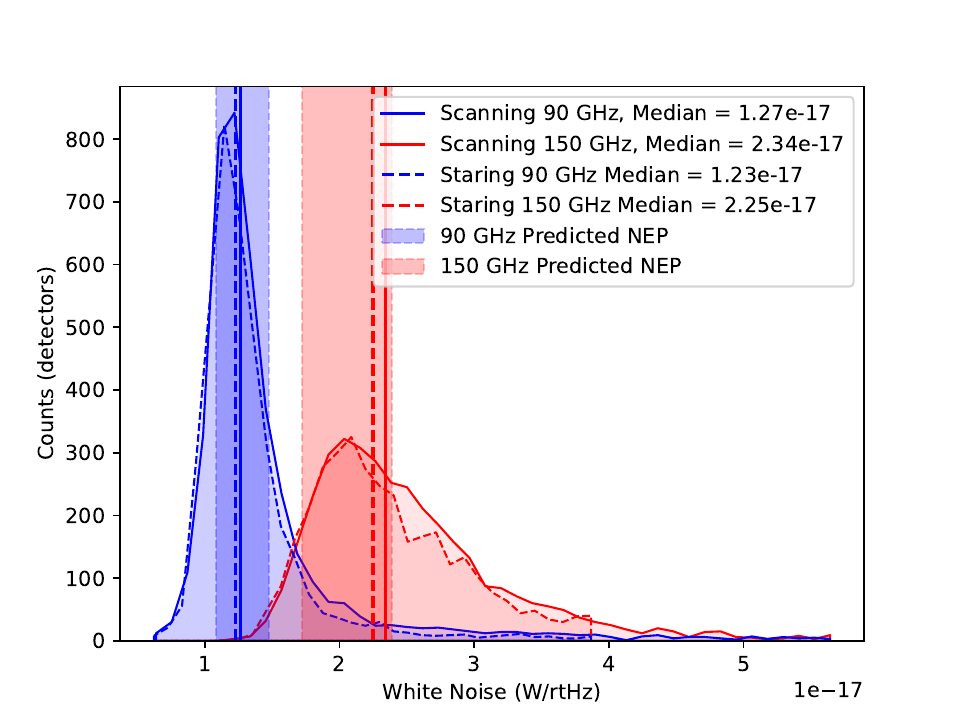}   
    \caption{White noise levels of our TODs with OTs capped at 4K. The solid line is when scanning and the dashed is when staring. Blue is the 90 GHz detectors and red is the 150 GHz detectors. The predicted bands are the expected noise for the dark configuration from BoloCalc.
}
    \label{fig:wn}
\end{figure}

The predicted noise equivalent power (NEP) ranges shown in Figure \ref{fig:wn} were calculated using BoloCalc \cite{Hill_2018}, a software package which estimates detector sensitivity, configured to mimic our dark configuration.
BoloCalc was also used in the forecasting done in \citenum{SO_Goals}.
This was done by using the same parameters as [\citenum{sierra2024simons}], but with all optics warmer than 4K excluded and the sky temperature set to 4K.
These NETs include contributions from photon noise, phonon noise, and readout noise.
It is important to note that this calculation relies on our inputs to BoloCalc being accurate and well understood,
but given the large number of inputs there is likely a relatively large error.

\begin{figure}[h]
    \centering
    \begin{subfigure}{.45\textwidth}
        \includegraphics[width=\textwidth]{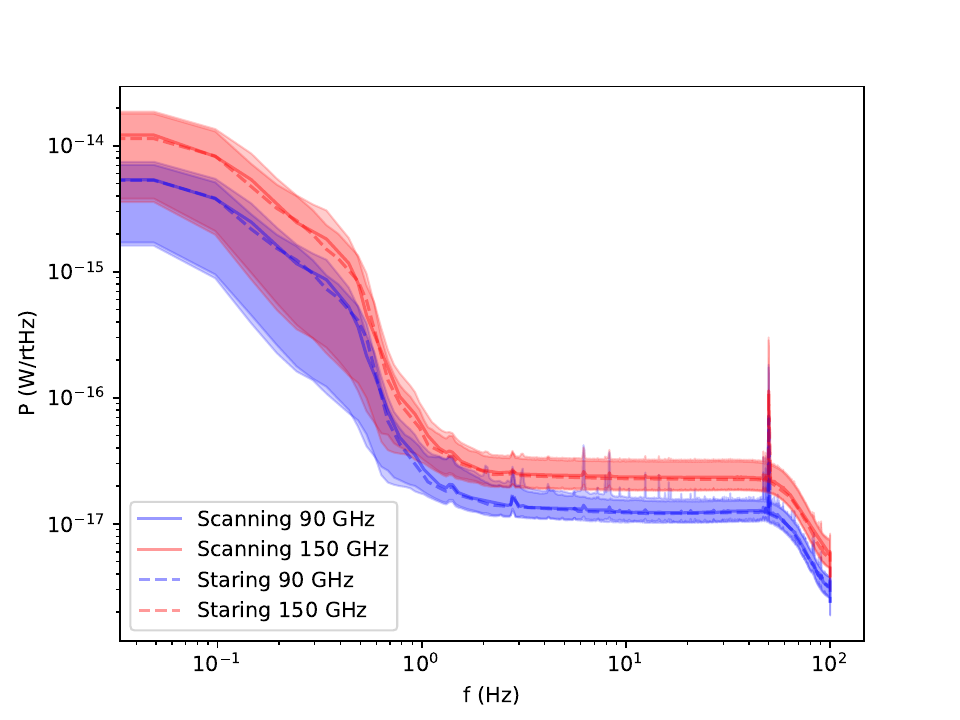}
        \caption{Before common-mode subtraction}
        \label{fig:ASD}
    \end{subfigure}
    \begin{subfigure}{.45\textwidth}
        \includegraphics[width=\textwidth]{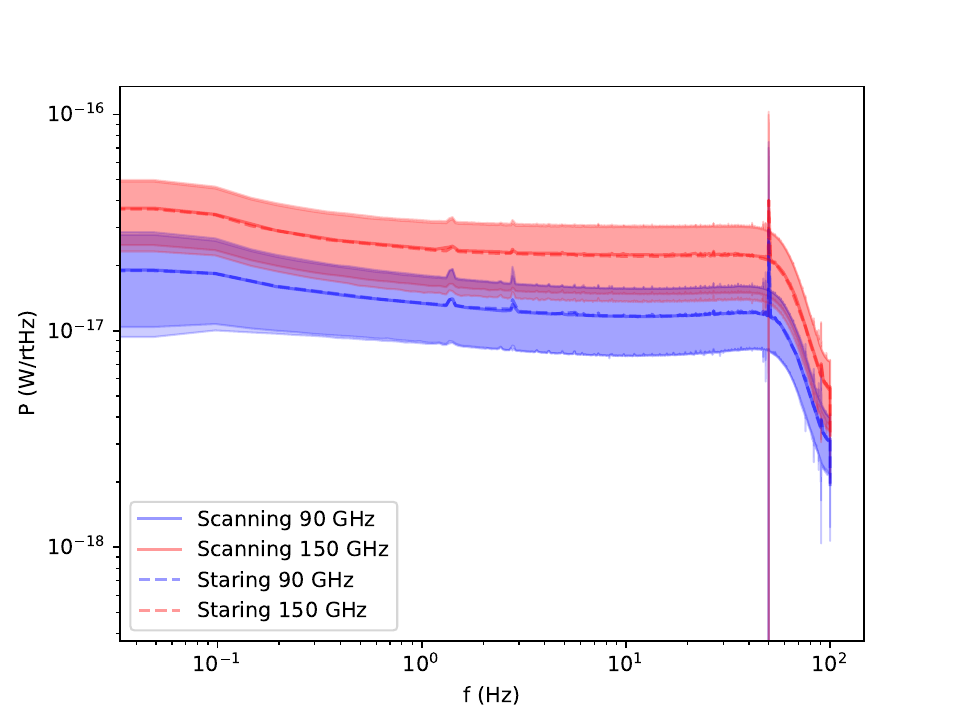}
        \caption{After common-mode subtraction}
        \label{fig:ASD_cmsub}
    \end{subfigure}
    \caption{The ASD of all detectors before and after common-mode subtraction The solid line is the median of all detectors when scanning and the dashed line is the median of all detectors when staring. With the contour being the 1 $\sigma$ level of these distributions. Blue is the 90 GHz detectors and red is the 150 GHz detectors. Note that the solid and dashed lines sit on top or very close to each other across the entire frequency range. The spikes in the ASD before removing the common mode are mostly sourced from an issue with a single UFM. The remaining bumps after common-mode subtraction are 50 Hz (AC power), 1.4 Hz (pulse tube operating frequency), and 2.8 Hz (harmonic of the 1.4 Hz bump) and coupled to the detectors in complex ways that are not common mode.}
\end{figure}

\subsection{Stare Noise}

Here we look at data taken while the telescope was stationary (``stare scan") as a baseline for comparison when considering noise while scanning (see Section \ref{noise_comp}).
This data was taken over the course of a single night ($\sim$eight hours) with all six UFMs.
This includes data from $\sim 8000$ biased detectors after cuts.

Our white noise levels (Figure \ref{fig:wn}) are tightly distributed, with the medians falling within the expected range.
The white noise distribution of 150 GHz detectors has a high tail that skews the median close to the edge of our predicted range, but the peak of the distribution is closer to the center of the predicted range.
Looking at the ASDs of our detectors, we can get an idea of what other features appear in our noise (see Figure \ref{fig:ASD}).
The sharp peak at 50 Hz is the AC frequency of power in Chile, and the slight bump around 2.8 Hz is a resonance of the operating frequency of the pulse tube coolers used by the LAT.
Of particular note here is that the low-frequency noise seems to be well-modeled by the common mode discussed in \ref{cm}, with a $\mathcal{O}(100)$ reduction in the low-frequency portion of the ASD after common-mode subtracting (compare Figures \ref{fig:ASD} and \ref{fig:ASD_cmsub}).
After removing the common mode (see Figure \ref{fig:ASD_cmsub},) we see that there is an additional peak at 1.4 Hz; this is the operating frequency of the pulse tube coolers.

\subsection{Scanning Noise} \label{scan}

Following stare tests, we conducted noise characterization tests of the detectors taken during a week of constant elevation scans (CESs).
This is the data used to produce the maps in Section \ref{maps} and includes data from $\sim 8000$ biased detectors after cuts.
Because the LAT was undergoing construction during the day, we only took scans at night; thus, the data volume was only $\sim 50$ hours.
The scans used were taken from a schedule that mimics the expected scan strategy for LAT on-sky observations.
All our CESs had a scan speed of 1 $^\circ/$s, an average turnaround acceleration of 1 $^\circ/\mathrm{s}^2$, and an azimuth throw of $60 ^\circ$.

Once again our white noise levels (Figure \ref{fig:wn}) are relatively tightly distributed, with the medians within the expected range.
The 150 GHz detectors display the same skewed distribution as when staring, with the median again falling closer to the edge of the predicted range than the peak.
The ASDs (Figure \ref{fig:ASD}) have the same 50 Hz and 2.8 Hz lines as the stare data, but here we see several additional high-frequency peaks.
We found that the cause of these peaks was a component of a single UFM (Mv11) erroneously dumping power in the presence of interruptions in the timing signal used by SMuRF; this issue has since been fixed.
Just as with the stare data, once the common mode is subtracted we see a $\mathcal{O}(100)$ reduction in the low-frequency noise (see Figure \ref{fig:ASD_cmsub}).
The common-mode subtraction is also able to remove the erroneous peaks associated with Mv11.

\subsection{Noise Comparison} \label{noise_comp}
While we expect that the noise during CESs may be worse than stare data due to the more challenging vibrational environment,
significantly worse noise would point to an issue in our receiver or scan strategy that needs to be addressed.
Here we examine this by comparing the results from the previous two subsections; note that all the ratios here are taken as $\frac{stare}{scan}$, such that a value less than 1 implies the scanning noise is higher.
All ratios are taken on a per detector basis, so we are always comparing the performance of a given detector to itself.

\begin{figure}[h]
    \centering
    \begin{subfigure}{.32\textwidth}
        \includegraphics[trim={0 0 0 1.37cm},clip,width=\textwidth]{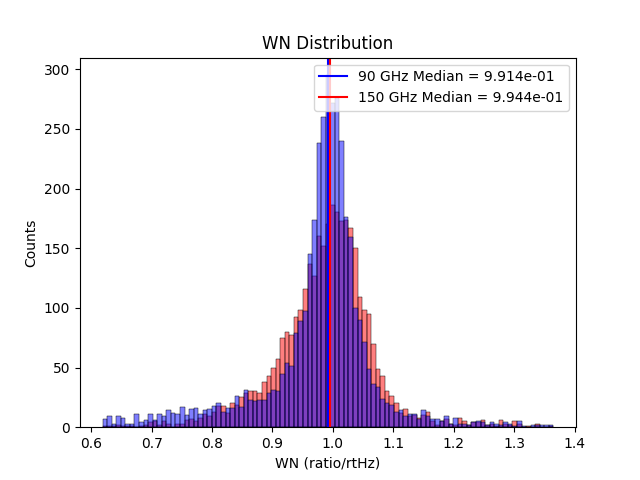}   
        \caption{Ratio of white noise levels.}
    \label{fig:ratio_wn}
    \end{subfigure}
    \begin{subfigure}{.32\textwidth}
        \includegraphics[trim={0 0 0 1.37cm},clip,width=\textwidth]{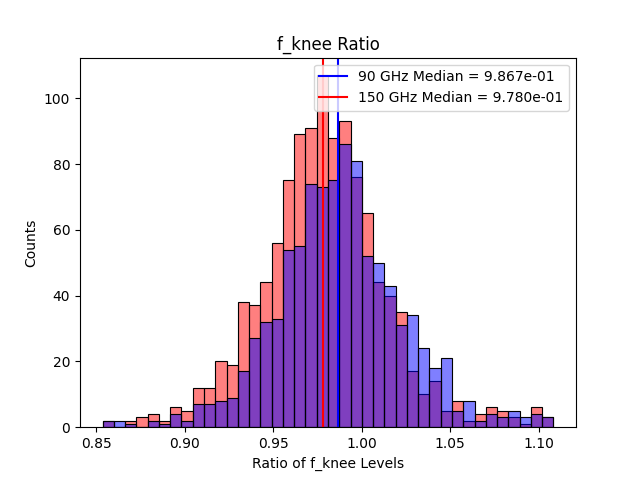}
        \caption{The ratio of $\frac{1}{f}$ knees.}
    \end{subfigure}
    \begin{subfigure}{.32\textwidth}
        \includegraphics[trim={0 0 0 1.37cm},clip,width=\textwidth]{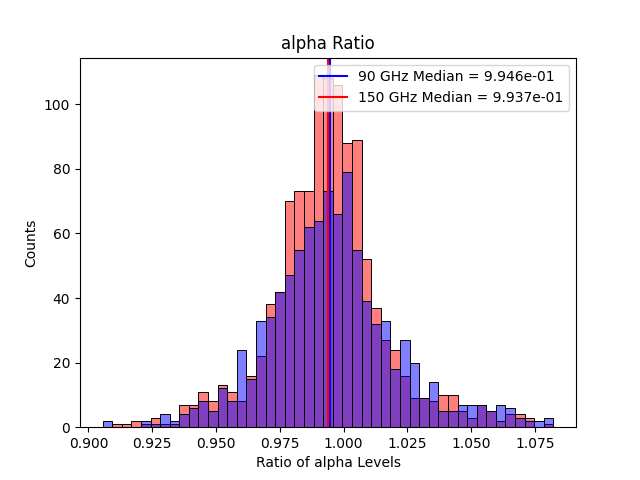}
        \caption{The ratio of $\frac{1}{f}$ indices.}
    \end{subfigure}
    \caption{The ratio of staring vs scanning noise parameters of all detectors. We see that both the white noise levels and $\frac{1}{f}$ parameters are statistically identical in both these states. Note that this is prior to common mode subtraction.}
    \label{fig:ratio_1f}
\end{figure}

\begin{figure}[h]
    \centering
    \begin{subfigure}{.45\textwidth}
        \includegraphics[trim={0 0 0 1.3cm},clip,width=\textwidth]{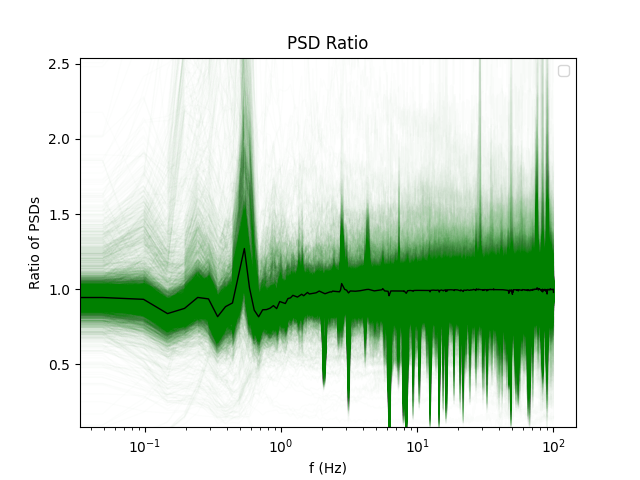}
        \caption{Before common-mode subtraction}
        \label{fig:ratio_ASD_pre}
    \end{subfigure}
    \begin{subfigure}{.45\textwidth}
        \includegraphics[trim={0 0 0 1.3cm},clip,width=\textwidth]{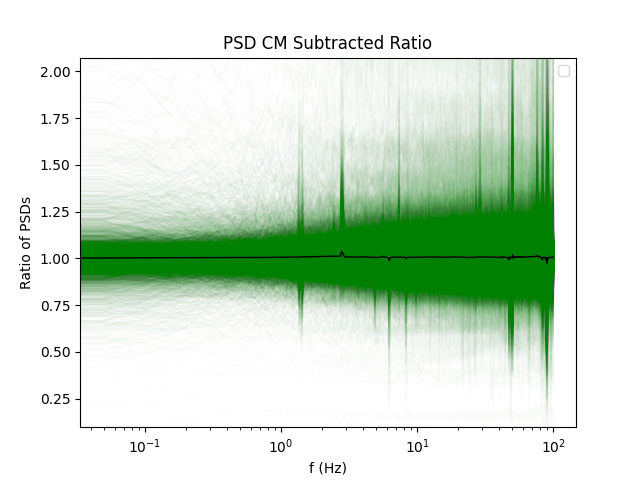}
        \caption{After common-mode subtraction}
        \label{fig:ratio_ASD_post}
    \end{subfigure}
    \caption{The ratio of staring vs scanning ASDs for all detectors. Black lines are the medians of all detectors and green lines are  $\sim$8000 individual detectors. This includes both 90 GHz and 150 GHz detectors.}
    \label{fig:ratio_ASD}
\end{figure}

%\begin{figure}[h]
%    \centering
%    \begin{subfigure}{.45\textwidth}
%        \includegraphics[trim={0 0 0 1.3cm},clip,width=\textwidth]{figures/tod_noise/ratio_PSD_sid.png}
%        \caption{Before common-mode subtraction}
%    \end{subfigure}
%    \begin{subfigure}{.45\textwidth}
%        \includegraphics[trim={0 0 0 1.3cm},clip,width=\textwidth]{figures/tod_noise/ratio_PSD_cmsub_sid.png}
%        \caption{After common-mode subtraction}
%    \end{subfigure}
%    \caption{The same data as Figure \ref{fig:ratio_ASD} but averaged by UFM. The spikes associated with Mv11 are well understood and have since been resolved.}
%    \label{fig:ratio_ASD_sid}
%\end{figure}

Looking at the ratio of the white noise levels (Figure \ref{fig:ratio_wn}), we find that the white noise while scanning is statistically identical to while staring.
Similarly, if we model the low frequency noise as $\frac{1}{f}$ noise (see equation \ref{eq:1f_noise}) and compare the $f_{knee}$s and indices (Figure \ref{fig:ratio_1f}), we see again that the staring and scanning noise are statistically identical.
Since we expect this low frequency noise to be dominated by variations in bath temperature in our dark configuration, this means that scanning has very low impact on our thermal stability.
As expected from the lower white noise and $\frac{1}{f}$ while staring, the median ratio of the ASDs (Figure \ref{fig:ratio_ASD_pre}) is $\leq 1$ almost everywhere, but there are some spikes with a ratio $> 1$ where the stare noise is higher than the scanning noise.
The spikes at frequencies above 1 Hz are to be associated with a relatively small number of detectors, as they do not shift the median away from $\sim1$, but the peak at $\sim0.6$ Hz is consistent across all detectors.
The origin of this $0.6$ Hz peak is still under investigation; the ratio of the common-mode-subtracted noise (Figure \ref{fig:ratio_ASD_post}) is $\sim1$ across the board, so it appears to be a difference in the common mode between these two states.
%Looking at these ratios by UFM (Fddigure \ref{fig:ratio_ASD_sid}), we see fairly consistent behavior across all of the UFMs\footnote{The peaks in Mv11 are understood and safe to ignore here, see Section \ref{scan} for details.}, with the common-mode-subtracted noise only disagreeing at the $\mathcal{O}(1\%)$ level.

\section{Dark Maps} \label{maps}
We use a simple mapmaker to test our our map noise and show that we are capable of handling the noise present in the dark configuration.
This is an important check to do, as these noise sources will also exist once on-sky, but it will be difficult to check this in the presence of strong atmospheric noise.
Additionally, mapmaking provides a simple visual check that the flagging described in Section \ref{tod_flag} is functioning as the phenomena that we flag create obvious features in the map (hot pixels, flat regions, etc.)

\begin{figure}[h]
    \centering
    \begin{subfigure}{.45\textwidth}
        \includegraphics[trim={0 0 2.95cm 0},clip,width=\textwidth]{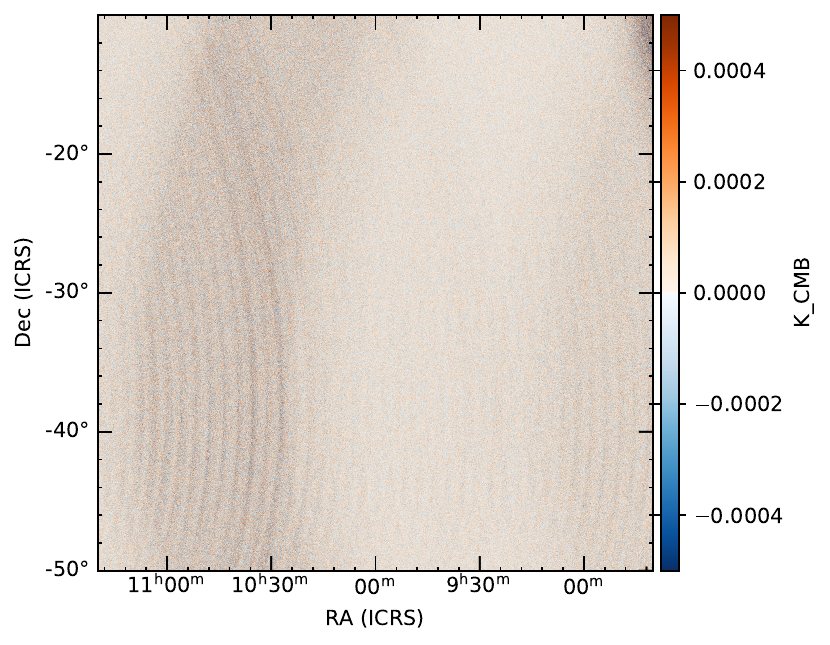}
        \caption{90 GHz map}
    \end{subfigure}
    \begin{subfigure}{.45\textwidth}
        \includegraphics[trim={0 0 2.95cm 0},clip,width=\textwidth]{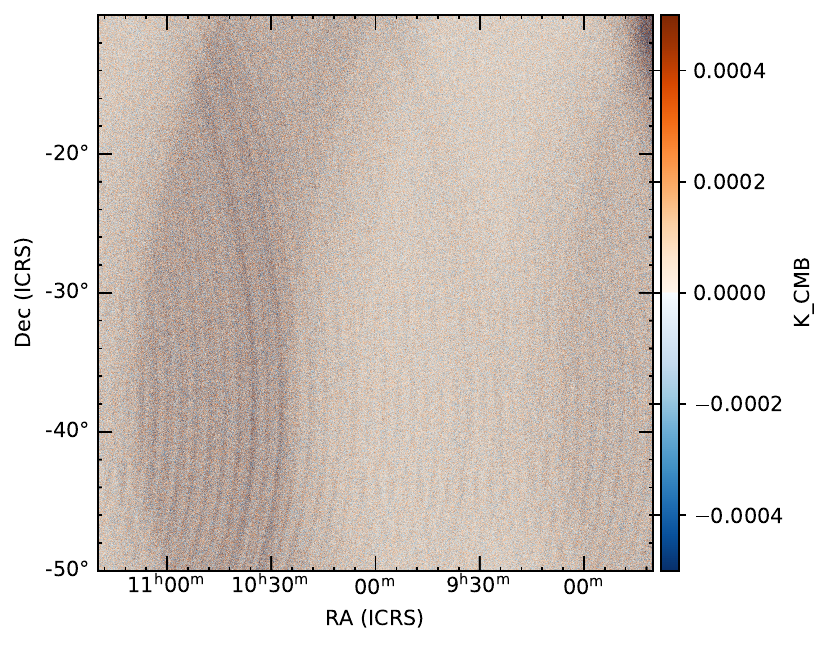}
        \caption{150 GHz map}
    \end{subfigure}
    \caption{A section of the dark maps made using equation \ref{eq:map_svd_sol}. The calculations done in Section \ref{map_noise} use a map that covers $\sim4$ times the area of those pictured here. The variations in noise levels across these maps are due to uneven coverage, with the arc-like structures corresponding to areas with poor overlap between scans. There is no obvious structure in the maps that point to significant unwanted data making it past our flagging.}
    \label{fig:maps}
\end{figure}

\subsection{Mapmaking Procedure} \label{mapmaking}
To turn our processed TODs into maps, we use the mapmaking equation

\begin{equation}
\label{eq:map}
    d = Pm + n \;,
\end{equation}
where $d$ represents the TODs, $P$ is the projection matrix that relates pixels to samples, $m$ is the map, and $n$ is the noise. Here we take a naive approach where we assume that the common mode as modeled in Section \ref{cm} completely describes the noise $n$, which gives us
\begin{equation}
\label{eq:map_svd}
    d = Pm + \mathrm{SVD}_{50}(d) \;,
\end{equation}
where $\mathrm{SVD}_{50}$ implies that we are taking the top 50 modes of the singular value decomposition as discussed in Section \ref{cm}. We can then solve equation \ref{eq:map_svd} for $m$, yielding
\begin{equation}
\label{eq:map_svd_sol}
    m = P^{T}N^{-1}(d - \mathrm{SVD}_{50}(d)) \;,
\end{equation}
where $N^{-1}$ is the inverse noise covariance from an uncorrelated noise model that simply has the inverse variance $\sigma^{-2}$ for each detector along the diagonal.
Note that we do not project flagged glitches into the map even after repairing them, since they would not contain any real signal if we were on-sky.

While this mapmaking scheme is not much more than a relatively naive filter+bin mapmaker,
it acts as a computationally cheap analog for the maximum-likelihood mapmaking used by ACT \cite{D_nner_2012} that will be the starting point for LAT mapmaking.
Maximum-likelihood mapmakers do not directly bin data into a map, rather they attempt to solve for the most likely (lowest $\chi^{2}$) map under the assumption that the noise is Gaussian and well described by $N^{-1}$.
This yields a map that is an unbiased estimate of the true map \cite{Cantalupo_2010}.
In this formalism, rather than subtract off our modeled noise such as in Equation \ref{eq:map_svd},
we use it to make an estimate of $N^{-1}$.
Once we have that, we can then solve for the maximum-likelihood map by solving the following linear equation:
\begin{equation}
\label{eq:map_ml}
    (P^{T}N^{-1}P)m = P^{T}N^{-1}d \;.
\end{equation}
Since the noise that we are subtracting in Equation \ref{eq:map_svd_sol} is a realization from the distribution described by $N^{-1}$,
equation \ref{eq:map_svd_sol} represents a biased way of solving for the the map.
Therefore, a map made with Equation \ref{eq:map_svd_sol} represents a lower bound on the accuracy of the maximum-likelihood map.
The noise covariance estimation in the ACT mapmaker \cite{Naess_2020} also is based on the SVD;
but employs a more sophisticated treatment, so what we have done here is just a naive approximation.

\subsection{Map Noise} \label{map_noise}
Since we have no signal in the dark configuration, we expect that with proper noise modeling the map noise should simply be white noise.
A simple test of this is to see if the root mean square (RMS) noise of the map agrees with what is expected from the measured NET of the TODs.
If the map's RMS noise is significantly higher, that implies that there is some non-white structure in the map skewing the RMS.

The RMS of a map with TODs containing only pure white noise is
\begin{equation}
\label{eq:n_white}
    n = n_{w}\sqrt{\dfrac{A_{pix}N_{pix}}{N_{det}t}} \;,
\end{equation}
where $n_{w}$ is the white noise level, $A_{pix}$ is the area of a pixel, $N_{pix}$ is the number of pixels in the map, $N_{det}$ is the number of detectors, and $t$ is the total time of the observations being mapmade.

Ignoring the non-white noise in the TODs, we can then compute the expected map RMS from TODs, $n_{tod}$, as
\begin{equation}
\label{eq:n_tod}
    n_{tod} = \overline{NET}_{det}\sqrt{\dfrac{A_{pix}N_{pix}}{N_{det}t}} \;,
\end{equation}
where $\overline{NET}_{det}$ is the average NET of a detector derived from the TODs.

For the map's RMS noise $n_{map}$, we weigh pixels by the number of hits (samples that have been binned into the pixel) giving us
\begin{equation}
\label{eq:n_map}
    n_{map} = \dfrac{\sum w_{i}m_{i}}{\sum w_{i}}\sqrt{A_{pix}} \;,
\end{equation}
where $w_{i}$ is the number of hits in the $i$th pixel and $m_{i}$ is the value of the $i$th pixel.

These values for our dark data are summarized in Table \ref{tab:map_noise} and show excellent agreement between the TOD and map values.
Note that these values were calibrated as described in Section \ref{tod_cal}, so while the self similarity is valid the actual values should not be interpretted.

\begin{table}[]
\centering
\begin{tabular}{|l|l|l|}
\hline
Band (GHz) & Map ($aW-arcmin$) & TOD ($aW-arcmin$) \\ \hline
90         & 3.86                 & 3.77 $\pm$ 0.57       \\ \hline
150        & 8.26                & 7.88 $\pm$ 1.76      \\ \hline
\end{tabular}
\vspace{5mm}
\caption{Map RMS noise values and expected values computed from the TODs. The uncertainties on the TOD values are computed from the spread on the distribution in Figure \ref{fig:wn}.}
\label{tab:map_noise}
\end{table}

\section{Conclusions}
The SO LAT will begin on-sky observations in 2025.
Ahead of first light, we were able to characterize detector noise in a dark configuration with caps at the 4K stage.
In this configuration, we see that we currently are within the expected range for white noise and that the non-white noise is largely common mode.
Comparisons of our noise while scanning to when stationary show that our nominal scan strategy has minimal impact on our noise.
The data processing techniques used for this work were able to successfully produce TODs usable for mapmaking, with the map noise in good agreement with what we expect from the TOD white noise levels.
While the flagging used for this work was able to cut or repair enough of the unwanted phenomena in our TODs to produce clean TODs, we currently flag a high portion of the data; work to address this issue prior to the LAT beginning on-sky observations is currently ongoing.

\section{Acknowledgements}
This work was funded by the Simons Foundation (Award \#457687, B.K.), the National Science Foundation (UEI GM1XX56LEP58), and the University of Pennsylvania.

% References
\bibliography{main} % bibliography data in report.bib
\bibliographystyle{spiebib} % makes bibtex use spiebib.bst

\end{document}